\documentclass[doublecol]{epl2}

\usepackage{graphicx,amssymb,color}

\title{Effects of electromagnetic waves on the electrical properties of contacts between grains}
\shorttitle{Effects of electromagnetic waves on...}
\author{S. Dorbolo \inst{1}, A. Merlen \inst{2}, M. Creyssels \inst{2}, N. Vandewalle \inst{1}, B. Castaing \inst{2}, and E. Falcon \inst{3}}
\shortauthor{S. Dorbolo \etal}
\institute{\inst{1} GRASP-Photop\^ole, Physics Department, University of Li\`ege, B-4000 Li\`ege, Belgium \\ 
\inst{2} Laboratoire de Physique, Ecole Normale Sup\'erieure de Lyon, 46 all\'ee d'Italie, 69 007 Lyon, France \\ 
\inst{3} Mati\`ere et Syst\`emes Complexes, Universit\'e Paris-Diderot -- Paris 7, CNRS, France }

\abstract{A DC electrical current is injected through a chain of metallic beads.  The electrical resistance of each bead-bead contacts is measured.  At low current, the distribution of these resistances is large and log-normal.  At high enough current, the resistance distribution becomes sharp and Gaussian due to the creation of microweldings between some beads.  The action of nearby electromagnetic waves (sparks) on the electrical conductivity of the chain is also studied. The spark effect is to lower the resistance values of the more resistive contacts, the best conductive ones remaining unaffected by the spark production. The spark is able to induce through the chain a current enough to create microweldings between some beads. This explains why the electrical resistance of a granular medium is so sensitive to the electromagnetic waves produced in its vicinity. }

 {Porous materials; granular materials}

\begin{document} 
   
\maketitle 

The electrical resistance of a granular assembly is very sensitive to a large variety of external perturbation.  The global electrical resistance can be indeed modified by a mechanical shock or stress, by a thermal dilatation, by aging \cite{jap}, by applying an electrical current \cite{falcon,vandenbrouk}, or by producing electromagnetic perturbation in its vicinity \cite{branly}.  The two last sources of perturbation are the most unexpected.  The relation between the voltage and the current injected through a granular material has been debated for a long time. It has been shown at the end of the XIXth century that a huge decrease of resistance occurs when a current is injected through metallic fillings \cite{onesti}.  The resistance drops over several order of magnitude when the current reaches a given threshold.  Almost during the same period, E. Branly discovered that an electromagnetic wave ({\it e.g.}, spark production in the air) is able to modify the electrical resistance of a granular heap at distance \cite{branly}.  This remarkable phenomenon is at the origin of the development of the wireless transmission. These problems have been recently revisited because the mechanisms are still not completely elucidated \cite{pls}.  Moreover the electrical properties could be a smart way to probe the internal structure of the mechanical arches through a granular pile.  When an electrical current is injected through a pile, it percolates according to the least resistance pathway which has a topological dimension of 1.  One of the challenges is to determine the exact role of the network compared to the role of one single contact with respect to the imposed perturbation.

\begin{figure}
\begin{center} ±
\includegraphics[width=6.5cm]{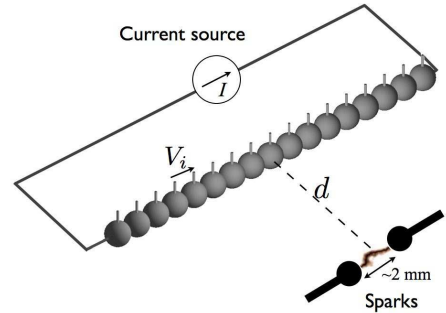}
\caption{Schematic view of the experimental setup.  The beads (8 mm of diameter) are placed into a groove and electrodes have been soldered on each bead. A current can be injected through the chain via the beads located at the extremities. Sparks can be also produced between two electrodes separated by 2 mm and located at a distance $d$ from the chain.} 
\end{center}
\end{figure}

\begin{figure*}
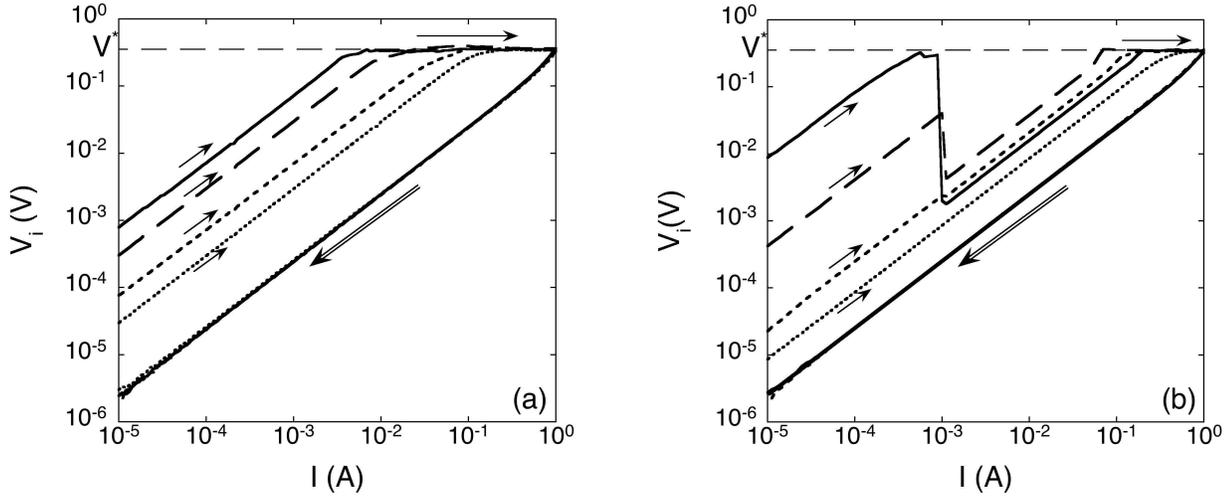
 
\includegraphics[width=8.5cm]{fig2a.epsf}
\includegraphics[width=8.5cm]{fig2b.epsf}
\caption{Typical current-voltage characteristics for 4 single contacts within the bead chain (different line styles).  (a)  Effect of a strong applied current : $I$ is first increased from 10 $\mu$A to 1 A, then is decreased to 10 $\mu$A.  (b) Effect of sparks: $I$ is first increased from 10 $\mu$A to 1 mA, then sparks are produced at 1 m from the chain.  The influence of the sparks is shown by the discontinuity in two curves.  Afterwards, $I$ is increased up to 1 A before being set back to 10 $\mu$A.  The single (resp. double) arrows denote an increase (resp. decrease) of the current.  All the 4 curves collapse on each other when the current is decreased. The horizontal dashed line shows the saturation voltage $V^*\simeq 0.4$ V (see text).} 
\end{figure*}

The relation between DC current and voltage has been described  for a one-dimensional (1D) chain of metallic beads \cite{falcon}, in a 2D configuration \cite{dorbolo2,mathieu} or in a 3D packing \cite{dorbolo1,creyssel}.  As reported in these works, the electrical properties of the bead assembly strongly depend on the electrical history of the granular pile.  The voltage is not univocally determined by the current because of irreversible processes such as microwelding occuring between the beads \cite{falcon}.   Some works have also revisited the influence of sparks on the electrical resistance of a 3D packing of lead beads (Branly effect) \cite{dorbolo3} and theoretically \cite{branly07}.

The aim of this paper is to study the effect of either a high current or an electromagnetic perturbation on the resistance of one single bead-bead contact.  By performing the experiments several times, a large number of contacts will be considered in order to establish the distribution of the resistances before and after the application of the perturbation.  The comparison of both situations will allow to determine the behavior of one single contact.

The experimental setup is shown in Fig. 1. Seventeen stainless steel beads (8 mm of diameter) are placed in a linear groove dug in a nylon block.  A screw allows to compress the chain of beads to about 100 N.   Electrodes are soldered on each bead.  Two more wires are soldered on the extreme beads in order to inject the current.  A stable current source (Keithley K2400) is used for this purpose.  A Keithley 2700 multimeter with a multi-channel card is used to determine the voltage between the successive beads.  That ensures a 4-wire measurement for the resistances of each contact.  Rhumkorf coils are used to produce sparks at different distances $d$ from the bead chain ($d=$ 0.1 m to 2.2 m).  The length of the sparks has been fixed to 2 mm, and its duration is roughly 500 ms.  This ensures that the sparks are produced as soon as the coils are switched.  In the original Branly's experiment, an antenna was fixed to the emitter (Rhumkorff coils) and to the granular medium.  The antenna allows for the amplification of the electromagnetic effects.   We decided not to use the antenna in order to prevent the masking of any effects due to the influence of a low power electromagnetic wave. Since the electrical properties of the granular materials are very sensitive to their electrical history, after each experiment, the system is reset:  the pressure on the beads is released and the beads are separated from each other.  Any possible microweldings between beads are then broken.  The measurements are performed up to 30 times, leading to roughly 500 measurements that ensure enough statistics.

The voltages $V_i$ between the bead number $i$ and $i+1$ are measured with respect to the injected current $I$ (see Fig. 1).  The voltage-current characteristics for 4 different bead-bead contacts are displayed in Fig. 2a with different line styles. The initial resistance $R_0$ for each contact is defined as the resistance at low current before any irreversible processes occur ({\it e.g.}, high current applied or electromagnetic wave production). $R_0$ are then extracted from each linear fit of the $V_i(I)$ curves between 10 $\mu$A $\leq I \leq 1$ mA.  The system is reset about 500 times to obtain enough statistics for the $R_0$ distribution.  As shown in Fig. 3a, this distribution is found to be very broad over 4 decades, and is well fitted by a log-normal distribution (see $\bullet$-symbols). A log-normal resistance distribution reflects the inhomogeneity of the oxide layer on the surface of each bead \cite{mathieu}.  The cumulative distribution function of $R_0$ is plotted in the Fig. 3b (solid line).  The mean value $\langle R_0 \rangle =38$ $\Omega$, and its standard deviation is about 155 $\Omega$. This log-normal distribution will be compared afterwards to the one obtained when a constraint is imposed on the chain. Two types of constraints can be imposed. Either a large current $\sim 1$ A is injected through the beads, or electrical sparks are produced in the vicinity of the chain. 
\begin{figure*}
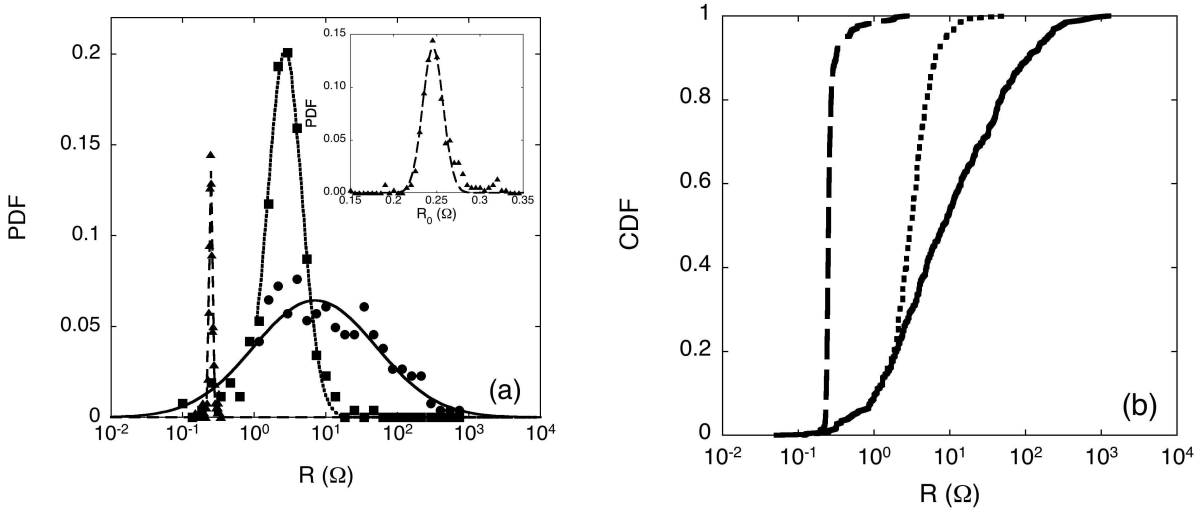
 
\includegraphics[width=8cm]{fig3aa.epsf}
\includegraphics[width=7.8cm]{fig3_b.epsf}
\caption{(a) Probability density functions (PDF) of the resistance $R_0$ at low current ($\bullet$), the resistance $R_{sc}$ after the injection of a current of 1 A  ($\blacktriangle$), and the resistance $R_{sp}$ after the production of sparks at 1 m from the chain  ($\blacksquare$).  The curves represent the log-normal distribution fit of $R_{0}$ ($-$) and $R_{sp}$ ($\cdots$) while a Gaussian is fitted to $R_{sc}$ ($--$).  The distribution of this latter has been plotted on a linear scale in the inset. (b) The cumulative distribution functions (CDF) of $R_{0}$ ($-$), $R_{sc}$ ($--$), and $R_{sp}$ ($\cdots$) are shown on the same semi-log plot.  The impact of sparks on the distribution is clearly visible between the $R_{0}$ ($-$) and $R_{sp}$ ($\cdots$)  CDF's: only the resistance values of the more resistive contacts have been decreased.}
\end{figure*}

First, we focus on the effect of a large applied current on the voltage-current characteristics. As shown in Fig. 2a, the current is first increased from 10 $\mu$A to 1 A (single arrows), and then decreased (double arrows).  A fine measurement of the voltage-current characteristic allows to extract several regimes.  As shown in Ref. \cite{falcon,mathieu}, when the current is increased, three different regimes occur: a linear one, followed by a nonlinear part, and then a saturation regime (see Fig. 2a).  When this latter regime is reached, the voltage between two successive beads can not exceed the saturation voltage $V^*\simeq 0.4$ V, and thus remains constant when $I$ is further increased \cite{falcon}.  This is due to an electro-thermal regulation within the contact.  In steady state conditions, the temperature of the contact can be expressed as \cite{falcon}
\begin{equation}
T_m^2-T_0^2=\frac{V^2}{4 L}
\end{equation}
where $V$ is the voltage across a contact, $T_{m}$ is the maximum temperature reached at the bead-bead contact, and $T_0=290 $K is the temperature far from the contact region, $L=2.45\times 10^{-8}$ V$^2/$K$^2$ being the Lorentz constant.   The contact geometry and the material characteristics do not appear in Eq.(1) because both the electrical resistivity, $\rho(T)$, and the thermal conductivity, $\lambda(T)$, are due to the conduction electrons, which leads to a linear temperature dependence $\lambda \rho = L T$, known as the Wiedemann-Franz law \cite{holm} (see Ref. \cite{falcon,greenwood} for details).  Moreover, the size of the micro-contact being much lower than the bead size and much larger than the electron mean free path (10 nm), Eq.(1) holds in a large range of contact size.  From Eq.(1), a low voltage near 0.4 V increases the temperature to about 1370 K at the center of the contact between two beads \cite{falcon}.  At such a high temperature, the micro-contacts between beads melt leading to microweldings. Since the temperature cannot exceed the melting value, it forces the voltage, from Eq. (1), to be a constant even when $I$ is further increased, leading to a decrease in the resistance $R_{sc}=V^*/I$ (see the plateau in Fig. 2a).  $R_{sc}$ denotes the value of the resistance of a bead-bead contact once the saturation regime is reached (that is as soon as a microwelding occurs). When the current $I$ is decreased, the resistance of the contact then remains equal to $R_{sc}$ (see double arrows in Fig. 2a). The distribution of the contact resistances $R_{sc}$ after the passage of a current of 1A is shown in Fig. 3a (see $\blacktriangle$-symbols).  The distribution is found to be very narrow, and is roughly fitted by a Gaussian (see dashed line in Fig.3a and in the inset).  The mean value is $\langle R_{sc} \rangle =$0.29 $\Omega$.  Since the distribution is very narrow, all the contacts can be viewed as electrically equivalent.  The large current has generated microweldings between some beads and has thus erased their initially high resistive values.  A finer description of the distribution is shown in the Fig. 3b by looking at the cumulative distribution function (CDF) of $R_{sc}$.  The distribution appears slightly asymmetrical due to a limitation at low resistance (the conductivity of the steel being finite).  

The intensity of the current has a different effect according to the initial resistance value of a single contact.  A current is qualified as large once a single contact voltage reaches the saturation regime (see Fig. 2a). Once such a large current is injected through the chain, the bead-bead contact resistances can be split into three groups: (i) the highly resistive contacts (with initial resistances $R_0 >> V^*/I^*$) become much more conducting (their resistances drop to $R_{sc} \approx V^*/I^*$), (ii) those with initial resistances $R_0 << V^*/I^*$ are not modified by the current, and (iii) the contacts with intermediate initial resistance are more or less affected by the current.

Let us now focus on the influence of sparks on the electrical properties of the chain.  Figure 2b shows typical voltage--current characteristics of 4 different bead-bead contacts with very different initial resistances $R_0$ (see different style lines). A current is first applied through the chain, and is increased from 10 $\mu$A to 1 mA (single arrows).   Then, sparks are produced using the Rhumkorff coils at a given distance $d$ from the chain of beads.  This leads to a voltage drop of some contacts, the corresponding $V_i$ - $I$ curves being thus discontinuous. Then, the current is further increased from 1 mA to 1 A.  The contact resistances measured just after the spark production are named $R_{sp}$.  Finally, the current is decreased back to 10 $\mu$A (double arrow).  The system is reset several times to obtain enough statistics for $R_{sp}$.
\begin{figure} 
\includegraphics[width=8.5cm]{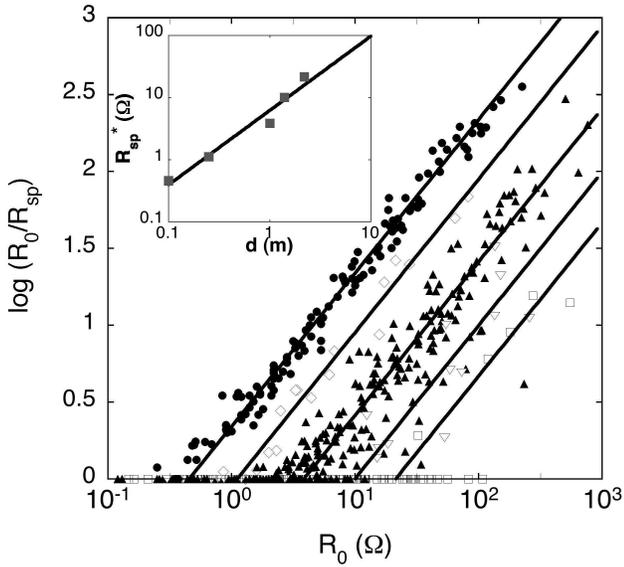}
\caption{Logarithm of the ratio between the resistances after, $R_{sp}$, and before, $R_0$, the spark production (at various distances $d$ from the chain) as a function of $R_0$. $d=$ 0.1 ($\bullet$), 0.25 ($\Diamond$), 1 ($\blacktriangle$), 1.4 ($\bigtriangledown$) and 2.2 ($\Box$) m.  Linear fits of the non-zero values of log($R_0/R_{sp}$) is displayed ($-$). Inset: $R^*_{sp}$ as a function of $d$.  Power law fit $\propto d^{1.2}$ is shown ($-$) as a guide for the eyes.} 
\end{figure}
Two kinds of behaviors appear. For inital high resistive contacts (large $R_0$), the sparks provoke a sharp voltage drop that betrays a drop of resistance (see the solid and dashed lines in Fig. 2b). The lower contact resistances remain unaffected by the spark production (see the dotted lines in Fig. 2b). The contact resistances that are influenced by the sparks drop to roughly the same value denoted $R^*_{sp}\simeq $ 2 - 4 $\Omega$ (see Fig. 2b) and remain stable up to saturation. Note that the only resistances much larger than $R^*_{sp}$ are affected by the sparks. The distribution of the resistance after sparks, $R_{sp}$, is displayed in Fig.3a (see $\blacksquare$-symbols).  This distribution is narrower than the initial one, but remains larger than the distribution of the resistance after the passage of a 1 A current.  The Fig.3b shows the cumulative distribution function of $R_{sp}$ well fitted by a log-normal (dotted line). Note that the CDF for $R_{sp}$ for values less than $R^*_{sp}$ remains the same as the $R_0$ one (see the identical part of the solid and dotted lines in Fig.3b).  The arithmetic mean of $R_{sp}$ is equal to 3.8 $\Omega$, about 10 times less than $\langle R_{0} \rangle$.

Let us now sum up all of our results. The injection of a large current through the chain has the same qualitative effect on the chain conductivity as that by the production of sparks in its vicinity.  This means that sparks can induce a current within the chain enough to create microwelding between some beads.  One can estimate the current induced by the electromagnetic waves as $I_{ind}\sim V^*/R^*_{sp} \sim 0.1$ A.  From Fig. 2a, such current of 0.1 A is indeed enough to generate microwelding between some beads. 

The role of the distance $d$ between the spark emitter and the beads on the chain conductivity is now examined. Several experiments are performed for different distances.  Figure 4 shows the logarithm of the ratio between the resistance after sparks, $R_{sp}$, and the initial resistance, $R_0$, as a function of the logarithm of $R_0$.  Thus, when the resistance is not affected by sparks, $\log(R_0/R_{sp})$ equals zero.  As said above, $R^*_{sp}(d)$ is the lowest resistance that sparks are able to change. Thus, for a fixed $d$, $\log(R_0/R_{sp})$ becomes non-zero only for initial resistances $R_0 > R^*_{sp}(d)$ (see Fig. 4).  $R^*_{sp}(d)$ are then extracted from the linear fits of the non-zero data in Fig. 4. The intersections of the solid lines with the x-axis give the values of $R^*_{sp}(d)$.  The inset of Fig. 4 show the log-log plot of $R^*_{sp}$ as a function of the distance $d$.  $R^*_{sp}$ roughly shows a power law dependence on $d$ with an exponent of 1.2.  As mentioned above, one can estimate the order of magnitude of the current induced by the sparks within the beads as $I_{ind}(d)\sim V^*/R^*_{sp}(d) \propto 1/d^{1.2}$. This seems to be close to the $1/d$ power-law expected for the electromagnetic waves in far field. However, since the bandwidth frequency of the emitted waves is unknown, this distance dependence deserves further works.  When the distance increases from $d=$ 0.1 m to 2.2 m, $I_{ind}$ is found to decrease from 0.87 A to 0.02 A, which remains large enough to produce microweldings between beads.

The infuence of either a high DC current or an electromagnetic perturbation on the electrical properties of a chain of beads has been studied.   The distribution of the bead-bead resistances before the perturbation is a log-normal over 4 decades.  Applying a high current transforms this distribution to a narrow Gaussian owing to the creation of microwelding between the contacts. When sparks are produced in the chain vicinity, only bead-bead resistances larger than a threshold value $R_{sp}^*$ are affected by the electromagnetic waves. $R_{sp}^*$ is roughly proportional to the distance between the chain and the spark emitter. Spark emission acts as a DC current which intensity inversely depends on the distance between the sparks and the contact.  This induced current is enough to create microweldings between some contacts. Generally, a granular packing has a huge contact number. Highly resistive contacts are thus likely.  Since only the largest resistances are influenced by sparks, this explains why the conductivity of a granular network is so sensitive to the action of nearby electromagnetic waves.

\acknowledgments
SD thanks FNRS for financial support and gratefully acknowledges the hospitality of the Physics Laboratory at ENS-Lyon. Part of this work was supported by the French Ministry of Research under Grant ACI 2001.  Dr. A.V. Orpe (Clark Univ.) is thanked for his comments.  The authors would like to thank C. Laroche for his helpful comments in the elaboration of the experimental set-up.

\end{document}